\documentclass[newstyle,twocolumn,proceedings]{rmaa}
\usepackage{rmaacite}

\title{Confidence limits of SN, Kinetic Energy and chemical yields in evolutionary synthesis models}

\altaffiltext{1}{Observatoire Midi-Pyr\'en\'ees \& CESR, Toulouse, France}
\altaffiltext{2}{Instituto de Astronom\'{\i}a, UNAM, M\'exico D.F.}

\author{M.~Cervi\~no\altaffilmark{1}
        and V. Luridiana\altaffilmark{2}}

\fulladdresses{
\item M. Cervi\~no: Observatoire Midi-Pyr\'en\'ees, 14 avenue Edouard Belin, 
31400 Toulouse, France (mcervino@obs-mip.fr).
\item V. Luridiana: Instituto de Astronom\'\i a,
  UNAM, Apdo. Postal 70-264, 04510 M\'exico D.F., Mexico 
  (vale@astroscu.unam.mx).}

\shortauthor{Cervi\~no \& Luridiana}
\shorttitle{Confidence Limitis: SNr, E$_k$ and $^{14}$N/$^{12}$C}

\keywords{Galaxies: evolution}

\abstract{% When evolutionary synthesis models take into account the
  stochastic nature of the IMF together with the discrete number of stars
  in real stellar clusters, typical output turns to dispersion band (where
  real data can be placed) instead of narrow lines.  We present here a
  qualitative analysis of such dispersion in the SN rate, the kinetic
  energy and the $^{14}$N/$^{12}$C ratio for different amounts of mass
  transformed into stars.}

\resumen{Cuando los modelos de s\'\i ntesis evolutiva tienen en cuenta la
  naturaleza estoc\'astica de la IMF y la discretizaci\'on del n\'umero de
  estrellas en c\'umulos reales, los resultados t\'\i picos de los modelos
  pasan de l\'\i neas infinitamente delgadas a bandas de dispersi\'on. En
  este trabajo presentamos un estudio cualitativo de dicha dispersi\'on en
  la tasa de SN, la energ\'\i a cin\'etica y el cociente $^{14}$N/$^{12}$C
  para c\'umulos con diferente cantidad de gas transformado en estrellas.}

\listofauthors{M.~Cervi\~no \& V.~Luridiana}
\indexauthor{Cervi\~no, M.}
\indexauthor{Luridiana, V.}
\begin{document}

\maketitle

\section{Motivation}

The influence of the stochasticity of the Initial Mass function (IMF)
on some outputs of evolutionary synthesis models has been studied in a
preliminary paper \cite{CLC00}. In that
paper, we showed that the resulting outputs present a dispersion which
depends on the amount of gas transformed into stars.

Here we investigate such dispersion for the Supernova rate (SNr), the
kinetic energy and the $^{14}$N/$^{12}$C ratio resulting from the evolution
of a stellar cluster. We have used the code presented in \scite{CMH94}
which performs a Montecarlo approximation to the IMF instead of the usual
analytical approximation. The models have been computed for a Salpeter IMF
slope, an upper mass limit equal to 120 M$_\odot$, and a lower mass limit
of 2 M$_\odot$, solar metallicity with standard mass loss rates
evolutionary tracks \cite{Sch92} and an Instantaneous Burst of star
formation. We have computed 500 Montecarlo simulations for a cluster with a
mass of 10$^4$ M$_\odot$, 200 simulations for a 10$^5$ M$_\odot$ cluster,
and 100 simulations for a 10$^6$ M$_\odot$ cluster.

\section{SNr}

\begin{figure}[Htb]
\begin{center}
  \includegraphics[width=\columnwidth, angle=-90]{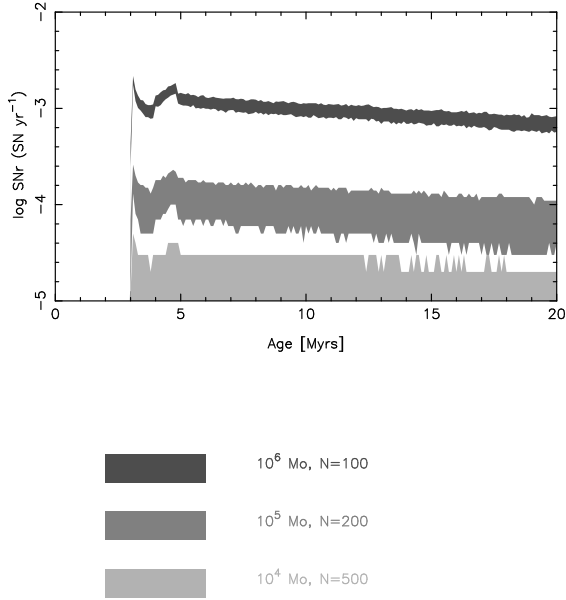}
\end{center}
  \caption{90\% confidence level for the SNr in function of the age of
  the stellar cluster. Different gray scale corresponds to different
  masses transformed into stars in the mass range 2 -- 120 M$_\odot$
  with a Salpeter IMF slope.}  \label{fig:fig1}
\end{figure}

The resulting dispersion with a 90\% confidence level for the SNr is shown
in the Fig. \ref{fig:fig1}. It can be shown that the dispersion in the SNr
increases with time since the number of SN decreases and the statistics
becomes lower \cite{Cetal01}.

\section{Kinetic energy and $^{14}$N/$^{12}$C ratio}

\begin{figure}
\begin{center}
  \includegraphics[width=\columnwidth, angle=-90]{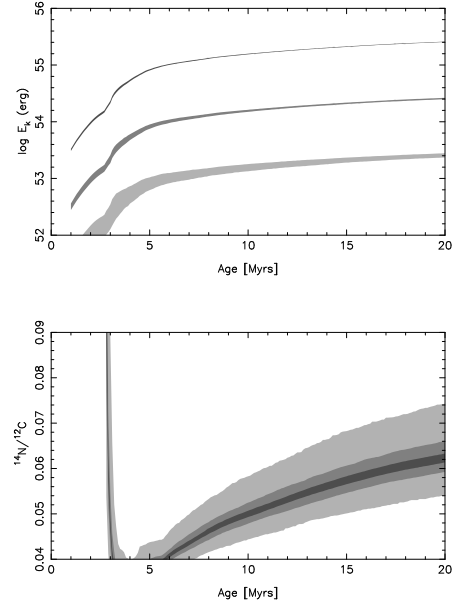}
\end{center}
  \caption{90\% confidence level for the kinetic energy produced by the
  cluster and $^{14}$N/$^{12}$C ratio in function of the age of the stellar
  cluster. Different gray scale corresponds to different masses transformed
  into stars in the mass range 2 -- 120 M$_\odot$ with a Salpeter IMF
  slope, following the color code of Fig.\ref{fig:fig1}.}  \label{fig:fig2}
\end{figure}

The resulting dispersion with a 90\% confidence level for the kinetic
energy produced by the cluster and $^{14}$N/$^{12}$C ratio is shown in 
Fig. \ref{fig:fig2}.

For the case of the kinetic energy, the dispersion comes from the
integration of Poissonian distributions over time. As the cluster evolves
the events become more and more numerous and the dispersion decreases.

The $^{14}$N/$^{12}$C ratio have a more complicated behavior. The
dispersion on the cumulative amount of individual $^{12}$C or $^{14}$N
decreases with time. But the dispersion of the $^{14}$N/$^{12}$C ratio
increases with time (like the dispersion in the SNr). From an
observational point of view, older clusters with the same amount of
gas transformed into stars will present a higher dispersion of this
ratio than younger ones.

\section{Discussion and Conclusions}

In general, the dispersion of synthetic observables of evolutionary
synthesis models due to the stochastic nature of the IMF is reduced
when the mass transformed into stars increases, but the relevance of the
dispersion is heavily dependent on the studied observable.

If the kinetic energy is the origin of temperature fluctuations, or, in
general, if temperature fluctuations are related with the stellar
population inside the nebula, the dispersion in the observed $t^2$ values
must increase for lower mass clusters. Such possible relation will be
investigated in \scite{LCB01}.

The dispersion in the $^{14}$N/$^{12}$C ratio becomes larger for older
clusters as far as the dispersion in the SNr increases. The extension of
this study to other metallicities and star formation histories remains to
be done and it can be useful to find a natural explanation of the observed
dispersion.

This study is included in a more general work abut these observables
presented in \scite{Cetal01} where a quantitative evaluation of the
dispersion is also presented.

{\small We want to acknoweledge the Instituto de Astronom\'\i a for some
facilities during this research. We}
\adjustfinalcols
\noindent {\small  want also acknowledge to the LOC
financial support.  MC is
supported by an ESA postdoctoral Fellowship.
}

\end{document}